\documentclass[fleqn,usenatbib]{mnras}

\usepackage{newtxtext,newtxmath}

\usepackage[T1]{fontenc}
\usepackage{booktabs}

\DeclareRobustCommand{\VAN}[3]{#2}
\let\VANthebibliography\thebibliography
\def\thebibliography{\DeclareRobustCommand{\VAN}[3]{##3}\VANthebibliography}

\usepackage{graphicx}	
\usepackage{amsmath}	
\usepackage{footnote}
\makesavenoteenv{tabular}  
\makesavenoteenv{table}

\newcommand{\hMpc}{{\ifmmode{\,h^{-1}{\rm Mpc}}\else{$h^{-1}$Mpc}\fi}}
\newcommand{\hkpc}{{\ifmmode{\,h^{-1}{\rm kpc}}\else{$h^{-1}$kpc}\fi}}
\newcommand{\hMsun}{{\ifmmode{\,h^{-1}{\rm {M_{\odot}}}}\else{$h^{-1}{\rm{M_{\odot}}}$}\fi}}

\newcommand{\Mstar}{{\ifmmode{\,M_{*}}\else{$M_{*}$}\fi}}
\newcommand{\Mhalo}{{\ifmmode{\,M_{\rm halo}}\else{$M_{\rm halo}$}\fi}}
\newcommand{\ltsima}{$\; \buildrel < \over \sim \;$}
\newcommand{\gtsima}{$\; \buildrel > \over \sim \;$}
\newcommand{\lsim}{\lower.5ex\hbox{\ltsima}}
\newcommand{\gsim}{\lower.5ex\hbox{\gtsima}}

\title[halo mass function]{How {\it much} do we know the halo mass function? Predictions beyond resolution.}

\author[W. Cui]{
Weiguang Cui,$^{1,2,3}$\thanks{E-mail: weiguang.cui@uam.es (WC)}\thanks{Talento-CM fellow}
\\
$^{1}$ Departamento de Física Teórica, M-8, Universidad Autónoma de Madrid, Cantoblanco 28049, Madrid, Spain\\
$^{2}$ Centro de Investigación Avanzada en Física Fundamental,(CIAFF), Universidad Aut\'{o}noma de Madrid, Cantoblanco, 28049 Madrid, Spain\\
$^{3}$ Institute for Astronomy, University of Edinburgh, Royal Observatory, Blackford Hill, Edinburgh EH9 3HJ, UK.\\
}

\date{Accepted XXX. Received YYY; in original form ZZZ}

\pubyear{\the\year{}}

\begin{document}
\label{firstpage}
\pagerange{\pageref{firstpage}--\pageref{lastpage}}
\maketitle

\begin{abstract}
As a common gravitation virialized object in the standard $\Lambda$CDM cosmology, dark matter halo connects from the large-scale structure all the way down to galaxy and star formation. However, as the nature of dark matter particles is still unclear, the smallest halo that can be formed in the universe is still unknown. Based on some simple assumptions, this paper uses the \textsc{hmf} package to investigate different halo functions used to quantify its number and mass distributions -- the halo mass function and the integrated/differential mass function (IMF/DMF) respectively. The halo mass in this study extends from the galaxy cluster to the dark matter particle mass at the GeV scale. 
Surprisingly, different fitting functions for the HMF are in remarkable agreement, a scatter within 2 orders of magnitude, down to dark matter particle mass, of which the halo mass spans about 80 orders of magnitude and the HMF covers over 100 orders of magnitude. The DMF reveals an interesting and consistent peak at $\sim 10^{13} \hMsun$, implying that galaxy groups contribute the most to the total matter mass. Furthermore, the effects of cosmology parameters on these halo functions are also examined with the most massive halos, or these halo functions at the massive halo mass end, more sensitive to them. Different behaviours of these halo functions due to the changes in cosmology parameters can be used to break the degeneracy between them.

\end{abstract}




\section{Introduction}

Cold Dark matter (CDM) \citep[see][for the latest comprehensive review]{Cirelli2024}, a pillar of our current standard cosmology model -- $\Lambda$CDM, collapses into highly dense regions due to gravity and initial density fluctuations. These high-density regions are referred to as dark matter halos. Under the hierarchical structure formation theory, smaller halos form first, and massive halos form later through both mergers and long-term slow accretion from their surrounding areas. As such, we will expect many more small halos compared to the massive ones at the current time in our Universe. The description of the halo distribution, i.e., the number of halos at different masses, is called the halo mass function (HMF). This is normally described as $\frac{d n}{d ln M}$, where $n$ is the number of halos at the particular mass $ln M$. To have a fair comparison between different models/simulations, this quantity is further normalised by the volume.

Though dark matter occupies about a quarter of the total Universe's energy content, it can not be directly observed and is difficult to measure in observation. Numerical simulations, on the other hand, provide theoretical guidance on how they look like. By sampling the whole density field with simulation particles, these N-body simulations can represent how the matter assembles through gravity attraction and, as such, the formation of dark matter and the large-scale structure \citep[see the Milliunum and MultiDark projects][for example]{Springel2006,Klypin2016}. Though the halos in the Universe formed by starting from small ones, the numerical simulations used to represent the universe by making the large-scale structure correct first with coarse resolutions, i.e. a small number of particles. There are several reasons behind this choice, (1) the simulation computation power requirement, in which an enormous number of particles is needed to cover many orders of spatial scales; (2) the unknown properties/nature of DM; and (3) resolving these very large-scale structures may not require extremely high-resolution simulations. Nevertheless, thanks to the boost of computation power as well as new efficient simulation codes, numerical simulations increase not only by resolution and volume, but also by modelling the baryon physics, which is a whole new research area that will be discussed briefly at the end of this paper. 

These gravitationally bound dark matter halos can be easily and visually found in N-body simulation, mathematically identifying them usually through either friend-of-friend (FOF) or spherically overdensity (SO) methods \citep[see][for a review]{Knebe2013}. Either method involves free parameters and different program packages (halo finders) have been developed in the community for this task \citep[see][for the comparisons]{Knebe2011}. Based on numerical simulations, the HMFs are calculated and fitted with various analytical functions \citep[see][for an incomplete collection]{Murray2013hmf}, which can be used for many different studies. Nevertheless, as shown in \cite{Murray2013hmf}, these fitting functions, especially those based on cosmological simulations rather than analytical formulae, have a limited halo mass range, which is simply due to the simulation resolutions -- dark matter halos require at least tens of particles to be identified, which is a free parameter of the halo finders. As such, even very recent, extra-scale suit of simulations, such as the Uchuu simulations \citep{Ishiyama2021}, can only cover about 9 orders of magnitude in halo mass. Note that this value is not achieved by running one single cosmological simulation, but by several simulations with different resolutions/box sizes. If dark matter is coming from the basic particles from the standard model with a mass, maybe $\sim 1$ GeV, there is no hope of running such a simulation, at least at the time of writing this article, with each simulation particle as a representation of the 'real' one. Nevertheless, if that could be done, we can expect solar-mass, even earth-mass, dark matter halos in the simulation. Thus, it is interesting to ask what are their properties and how they are distributed.

The pioneering work by \cite{Diemand2005} was the first to see Earth-mass dark-matter haloes at the very beginning of the structure formation ($z = 26$) with the simulation resolution reaching particle masses within their highest resolution region.
Thanks to \cite{Wang2020}, who used the zoomed-in technique to run a series of simulations to probe the dark matter halos in the void environment over 20 orders of magnitude all the way done to $z = 0$. This, again, is achieved by the zoomed-in simulation technique \citep[see more details in their paper][]{Wang2020}. The halo density profiles are confirmed to be universal over the entire halo mass range and are well described by a simple two-parameter fitting formula \citep{Navarro1997}. Furthermore, using the same set of simulations, \cite{Zheng2024} showed that an extended Press–Schechter formulae \citep[see][etc. respectively]{Press1974,Mo1996} can still fit the HMF in the void regions at different simulation resolutions. Though 20 orders of magnitude is a great number to be achieved with N-body simulations, it is nevertheless far from directly simulating dark matter particles, which may be at the GeV scale. However, given that the halo density profile is more-or-less unchanged from extreme low-mass halos to galaxy clusters, and the fitted HMF can still be well matched to the one from zoomed-in void simulation over many orders of magnitude in halo mass, one can naively speculate that the HMF fitting functions can be extended to these extremely low-mass halos, which maybe still match the cosmological simulations with infinite box size and resolution. If so, what can we learn/predict from this? 

In this paper, I adopt different HMF-fitting functions from the literature, which have been gathered and implemented in the \textsc{hmf} package\footnote{https://github.com/halomod/hmf}\citep{Murray2013hmf}, to conduct these experiments. Therefore, the \textsc{hmf} and other analytical calculations are first introduced in \autoref{sec:hmf}. All the results in \autoref{sec:results} are set at $z=0$: I first study the differences between different fitting functions in \autoref{subsec:hmf}; then the integrated and differential masses functions in halos in \autoref{subsec:idm}; lastly, the effects of cosmology parameters in \autoref{subsec:cosmo}. The conclusions with some discussions on the potential meanings/implementations and some possible influences from other processes are listed in \autoref{sec:conc}.

\section{Methods}\label{sec:hmf}
\subsection{The \textsc{hmf} and HM-fitting functions}

The HMFs are calculated by using the \textsc{hmf} package, which has been detailed in \cite{Murray2013hmf} for the usage (see also its online documentation at \url{https://hmf.readthedocs.io/en/latest/index.html}) and \cite{Murray2013} for its science aim, detailed algorithms and different applications. As such, I only briefly list these used fitting functions here, as well as detailed modifications and specified parameter usages in this study.

\subsubsection{HM-fitting functions}

\begin{table*}
\centering
\caption{The HMF fitting functions included in \textsc{hmf}. The name column indicates the name of the fitting function used in this paper.}
\label{tab:FF}
\makebox[\textwidth][c]{
\begin{tabular}{lc|lc} 
 \toprule
    \multicolumn{2}{c}{FOF} & \multicolumn{2}{c}{SO}\\
    \cmidrule(lr){1-2}\cmidrule(lr){3-4}
    Name & REF. & Name & REF.\\
    \cmidrule(lr){1-2} \cmidrule(lr){3-4}
    Jenkins & \cite{Jenkins2001} & PS$^*$ & \cite{Press1974} \\
    Warren & \cite{Warren2006}  & SMT & \cite{Sheth2001} \\
    Reed03 & \cite{Reed2003}  & Behroozi & \cite{Behroozi2013} \\
    Reed07 &  \cite{Reed2007}  & Watson    &  \cite{Watson2013}  \\
    Peacock&  \cite{Peacock2007}  & Tinker08  & \cite{Tinker2008}   \\
    Angulo &  \cite{Angulo2012}  & Tinker10  &  \cite{Tinker2010}  \\
    AnguloBound&  \cite{Angulo2012}  & Bocquet200cDMOnly & \cite{Bocquet2016}   \\
    Watson\_FOF&  \cite{Watson2013} & Bocquet200cHydro  &    \\
    Crocce  & \cite{Crocce2010}  & Bocquet500cDMOnly &   \\
    Courtin &  \cite{Courtin2011} & Bocquet500cHydro &   \\
    Bhattacharya  & \cite{Bhattacharya2011}  & - & -  \\
    Pillepich &\cite{Pillepich2010} & - & -  \\
    Manera  & \cite{Manera2013}  & - & -  \\
    Ishiyama&  \cite{Ishiyama2021} & - & -  \\
\bottomrule
\multicolumn{3}{l}{$^{*}$\footnotesize{Note this is an analytical formula, it is included in this role only to save some space.}} \\
\end{tabular}
 }
\end{table*}

To estimate the halo abundance, \cite{Press1974} (hereafter PS, see also \citet{Bond1991} for the extended PS formalism) established a basic paradigm based on the assumptions that (1) dark matter haloes form at peaks in the linear density field exceeding a predetermined threshold, $\delta_c = 1.686$ is the critical overdensity for spherical collapse; (2) the halo mass enclosed within a sphere of a radius R, with which the mass variance $\sigma (R)$ is linked to the integration's of the matter power spectrum $P(k)$ of the universe. As such, the HMF can be expressed as:
\begin{equation}
    \frac{dn}{d \ln{M}} = \frac{\rho_0}{M} f(\sigma) \left|\frac{d \ln{\sigma}}{d \ln{M}}\right|;
    \label{eq:1}
\end{equation}
with $\rho_0$ as the mean density and $f(\sigma)$ representing the functional form that defines a particular HMF fit, or $f(\sigma) = \sqrt{\frac{2}{\pi}}\frac{\delta_c}{\sigma} exp\left( \frac{\delta_c^2}{2 \sigma^2} \right)$ in the case of PS.
However, the PS analytical formalism doesn't provide a precise agreement to the N-body simulations \citep[see][for example for more discussion on this]{Jenkins2001, Sheth2001, White2002}. Subsequent studies on HMF adopted the same philosophical approach with the HMF described by \autoref{eq:1}, but using different functions for $f(\sigma)$ to characterise the HMF based on fitting the dark matter halo mass distribution from N-body simulations. 

In \autoref{tab:FF}, an incomplete list of HMF-fitting functions from the literature is presented. Note that only these have been included in the \textsc{hmf} listed in this table. Because this paper does not investigate the differences between these HMFs, I hope this sample is enough to clarify the point. These fitting functions are separated into two families: FOF (friend-of-fried) and SO (spherical overdensity), based on the halo-finding algorithms. The detailed formula for $f(\sigma)$, their fitting parameter values, as well as the limited application range of $\sigma$ are not shown in the table. It is recommended to \cite{Murray2013hmf} and these literature papers for such details. Both FOF and SO methods identified dark matter halos depend on the choices of free parameters: the linking length for FOF and the overdensity \footnote{This is normally with respect to the mean or critical overdensity of the Universe at corresponding redshift. Furthermore, \cite{Bryan1998} proposed a ``virial" overdensity (referred to as VIR), which has a value of about 198 instead of the commonly adopted 500 or 200 (referred to as 500c and 200c respectively). } for SO. This study adopted the widely used values: 0.2 for linking length and 500c/200c/VIR for overdensity. Because these fitting functions are proposed along with specified halo definitions, the study in this paper tries to keep them as it is instead of converting them into the same definition -- a feature included in \textsc{hmf}.

\subsubsection{The modification and parameter setting}

As mentioned in the introduction, it is interesting to know the smallest dark matter halos formed in the universe and how they are distributed. As such, the minimum halo mass is set to $10^{-60} \hMsun$ and the maximum halo mass is $10^{17} \hMsun$. Besides these assumptions in the introduction, it is necessary to go beyond the $\sigma$ ranges of these fitting functions. This extension yields an interesting agreement though it is not recommended, which will be presented in the result \autoref{sec:results}. Other than that, there are several extra cautions are included in this experiment. Because the HMF is going to be extended into particle mass scale, GeV from the massive galaxy clusters with a mass of $\sim 10^{16} \hMsun$, the accuracy in the \textsc{hmf} calculation needs to be taken into great care. Therefore, the code is updated to use long double instead of the standard float type of Numpy. Furthermore, $\sigma$ is determined by the matter power spectrum, which is also extended to extreme values: $\ln{k_{min}} = -30$ and $\ln{k_{max}} =100$. Lastly, the power spectrum is calculated by 
\begin{equation}
    P(k) = Ak^nT^2(k)
    \label{eq:2}
\end{equation}
where $T(k)$ is the transfer function, $A$ is the normalisation constant and $n$ is the spectral index. A is calculated by the cosmological parameter $\sigma_8$, the mass variance on a scale of $8 \hMpc$. Also, note here that the Planck 2015 cosmology parameters \citep{Planck2016} are used if it is not particularly mentioned in this paper. \textsc{hmf} provide different transfer functions to use. It is fixed to the \cite{Eisenstein1998} fitting function with BAO wiggles. However, the differences between these transfer functions are detailed in the \autoref{app:1}.


\subsection{The integrated and differential mass functions}

A simple question to ask is how much matter mass in the Universe is located in dark matter halos. Or will all dark matter particles collapse into a halo? This question can be answered by simply integrating the HMF. This will be called the Integrated mass function for halos (IMF) to distinguish the conventional terminology of HMF. As the HMFs are usually normalised by volume, the direct integration will result in the total mass within a certain volume, i.e. the mean density of the universe. Furthermore, if the integration starts from the most massive halos, we can expect to see its most significant influences at these low-mass halos. The other reason for this choice is most of these fitting functions are limited by halo mass $\gsim 10^8 \hMsun$ due to the simulation resolution. 
Another note here, also mentioned before, the maximum halo mass is set to $10^{17} \hMsun$, and there is a very tiny change in the integrated mass function with the starting maximum halo mass from $10^{30} \hMsun$. The reason is straightforward: the sharp turn in the HMF at $M_{halo} \approx 10^{14} \hMsun$ dramatically decreases the number of massive halos and their mass contributions. Lastly, the integration of the extended HMFs from \textsc{hmf} is done with the Scipy package, the Simpson function within its integration module. 

A furthermore interesting question is, at which halo mass range, do they make the highest contribution to the total matter mass in the universe? This is also an easy task as soon as the integrated mass function is derived or simply sums up the halo masses, instead of the number as the case in HMF, inside each mass bin. This mass abundance is called the differential mass function for halos (DMF) later on. Since the IMF provides the total mass ($\sum M(>M_{halo})$) at each $M_{halo}$, it is straightforward to have the DMF from the IMF. Similar to IMF, DMF also has volume normalisation, one can get its expression in fractions by simply dividing the mean density. To avoid any confusion later on, we refer to the HMF, IMF and DMF as the halo functions.

\section{results} \label{sec:results}

\subsection{The similarity of different HMF-fitting functions}\label{subsec:hmf}
\begin{figure*}
    \centering
    \includegraphics[width=\textwidth]{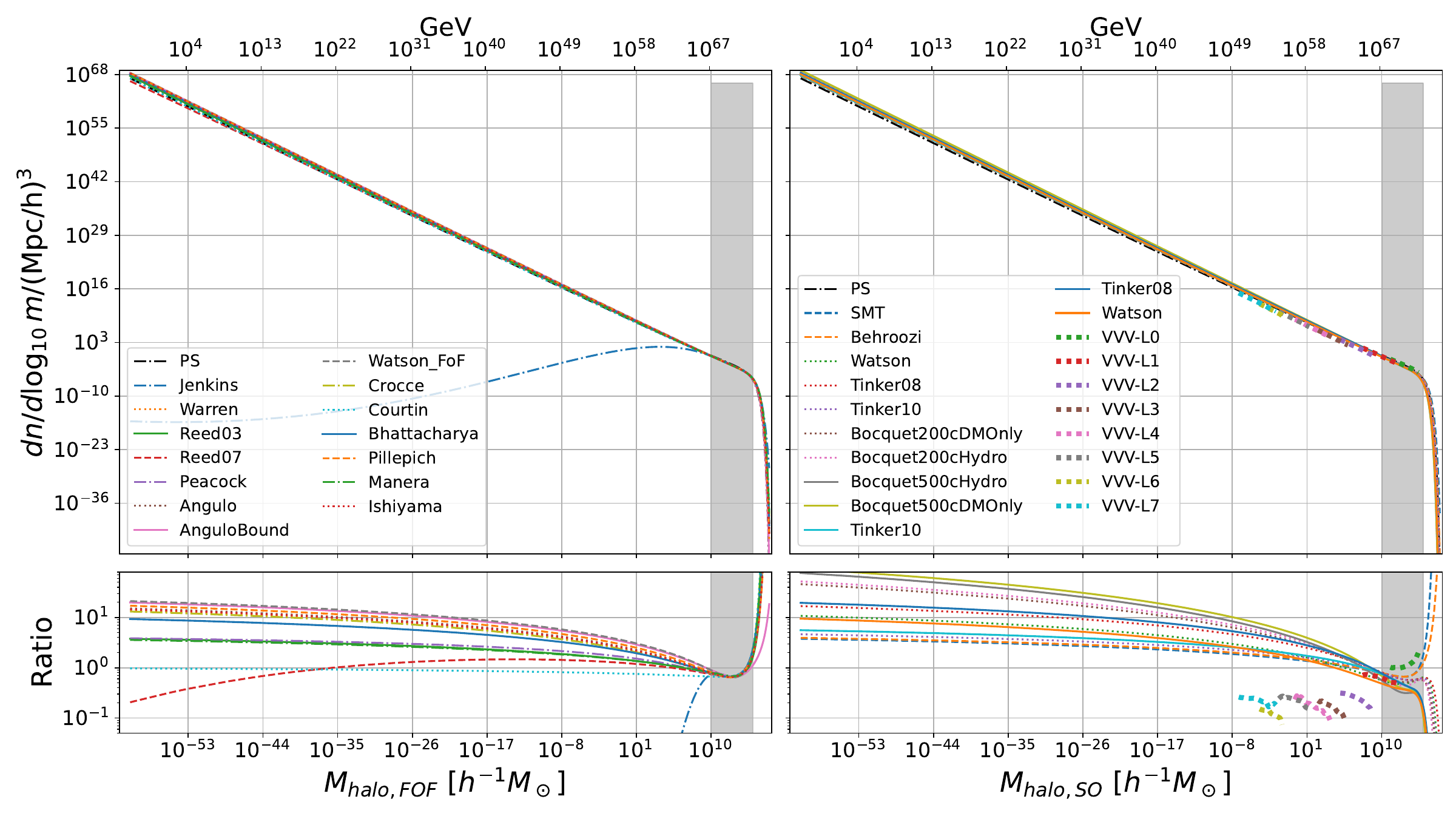}
    \caption{The HMF as a function of FOF halo mass (left panel) and SO halo mass (right panel). The different line styles and colours on the left panel mark different fitting functions as indicated in the legend. Different choices of overdensity parameters specify the line style on the right panel: dashed for VIR, dotted for 200c and solid for 500c. The HMFs from \citet{Zheng2024} are shown with thick dotted lines. The dash-dotted lines are reserved for the PS results, shown in both panels. Furthermore, the lower panels show a relative result compared to the PS HMF. The vertical grey region shows roughly the halo mass limitations ($10^{10} - 10^{15} \hMsun$) of these fitting functions based on N-body simulations. }
    \label{fig:hmf}
\end{figure*}
Besides the analytical formulae from \cite{Press1974} and \cite{Sheth2001}, all the other functions listed in \autoref{tab:FF} are fitting $f(\sigma)$ with numerical simulation results. Therefore the lower limits of these fitting functions are primarily determined by the simulation resolution. As such, there is no guarantee that they should agree beyond those limits, especially with the extension down to dark matter particle mass scale which is about 70 orders of magnitude lower. 

In \autoref{fig:hmf}, the different HMFs, separated into two families based on the halo definitions, are shown from $M_{halo} \approx 10^{-60} \hMsun$ to the massive galaxy clusters, $\sim 10^{17} \hMsun$. The analytical result from PS is shown in both panels and used as the reference line to highlight the relative differences at the lower panels. Besides Jenkins, very little difference shows up on the top panels due to the huge range this plot covers. Note that it is well-known that the PS model overpredicts the HMF at $M_{halo} \lesssim 10^{13} \hMsun$, but gives a sharp decrease at the high-mass end, $M_{halo} \gsim 10^{15} \hMsun$. Nevertheless, this won't affect the relative differences shown in \autoref{fig:hmf}, which we are mostly interested in. The agreements between different models look extremely good within the halo mass limits, demonstrated by shadow regions. The deviations increase down to about Earth's mass or even lower. Then, the relative differences become more or less stable until the dark matter particle mass scale. 

Again, besides Jenkins, it is surprising to see these relative differences are basically within 2 orders of magnitude with the extension covering about 80 orders of magnitude in halo mass and over 100 orders of magnitude in the HMF as indicated in the y-axes. Note that on the right panel, these results based on SO methods tend always to have higher values than the PS result. This implies that the SO method tends to enclose a smaller region than the FOF method, which results in an increased number of tiny halos. 

At the massive halo mass end, the difference becomes more violent. On one hand, the sharp drop in HMF, over 40 orders of magnitude decrease within a halo mass change of $\lesssim 2$ orders of magnitude, makes the prediction very sensitive; On the other hand, adopting the PS results makes this situation worse, which systematically lower than the FOF method result and higher than the SO method result (besides SMT and Behroozi, both of which use VIR overdensity instead of 500c and 200c). Nevertheless, this sensitivity can provide a strong constraint on the HMF, which can be used to constrain cosmology parameters, which will be detailed in the following sub-sections.

Lastly, the VVV simulation results are also included on the right panel for comparison. Agreed to the findings in \cite{Zheng2024}: the HMFs from the VVV simulations are lower compared to the other results, besides L0 and L1. This can be understood as the VVV simulations simulating the particular void regions of the universe, while all the others include different environments. There are slightly large differences compared to the ones claimed in \cite{Zheng2024}. This could be simply because the PS instead of their specified EPS formula is adopted. 

\subsection{The integrated and differential masses in dark matter halos: IMF and DMF} \label{subsec:idm}
\begin{figure*}
    \centering
    \includegraphics[width=\textwidth]{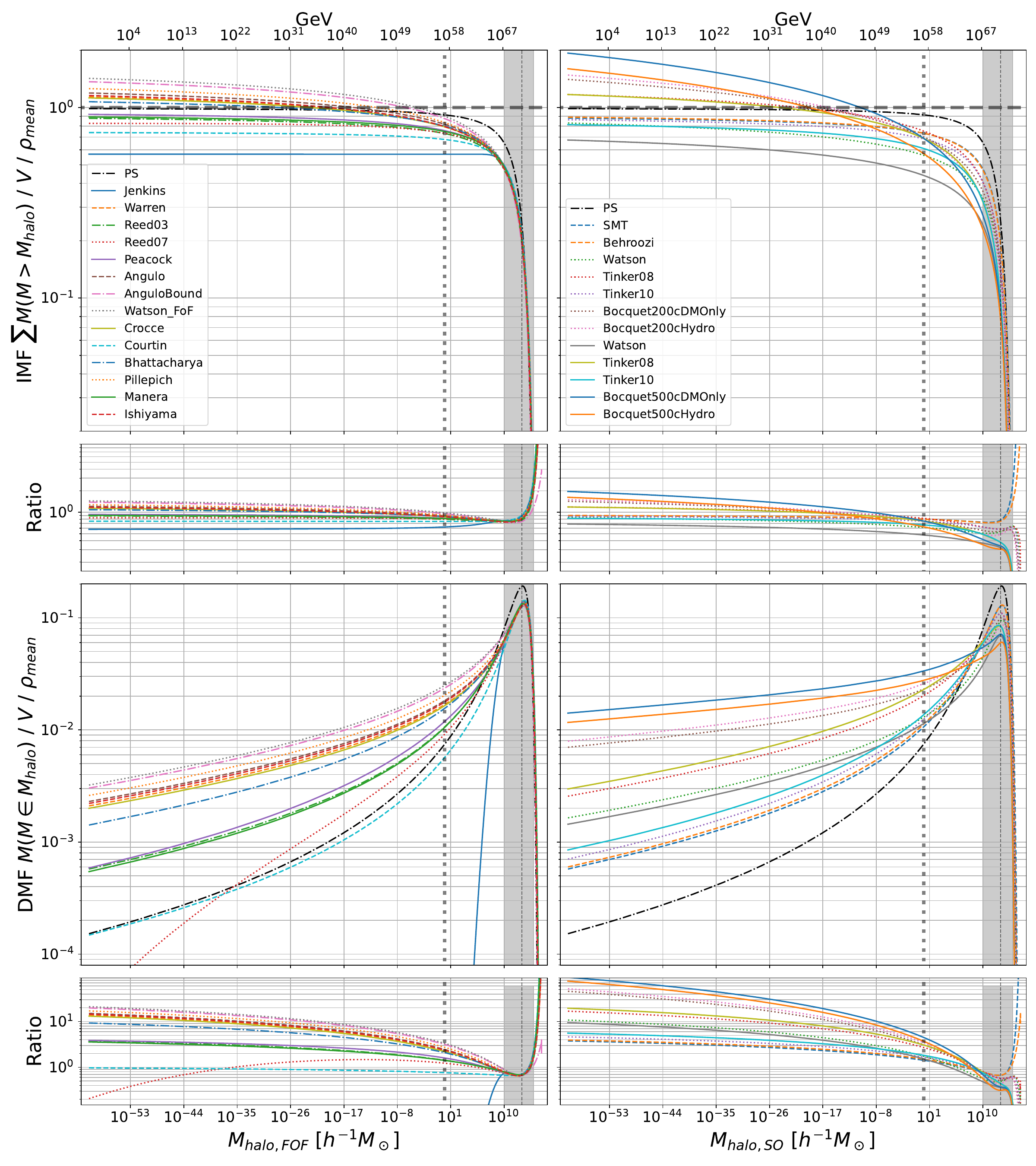}
    \caption{Similar to \autoref{fig:hmf}, the IMF and DMF are shown for FOF halos on the left-hand side panels and SO halos on the right-hand side panels, respectively. Note that the y-axes show the dimensionless fractions after the normalisation of the universe mean density, $\rho_{mean}$. Again the relative differences shown on these bottom panels are showing the ratio to the PS result and the different line styles for SO halos are for different overdensities as presented in \autoref{fig:hmf}. The shadow regions indicate these fitting function limits, the vertical think dotted lines mark the solar mass and the vertical thin dashed lines indicate $10^{13} \hMsun$, where these peak positions roughly are for the DMFs. Furthermore, the fraction of 1 in the IMF plots is highlighted by two horizontal thick dashed lines.}
    \label{fig:idmf}
\end{figure*}

The HMF describes the halo abundance at different masses, which provides a clear view of the halo distribution in number counts. While the IMF and DMF provide interesting views from a different point of view: their mass contributions. Though the halo numbers increase as power law as halo mass decreases, their masses in each mass bin do not. The reason is simply their masses decrease as well. By integrating the HMF from the most massive halos, the IMF indicates the mass fractions which are located within halos above a given halo mass. By extending the HMF down to dark matter particle mass, one would expect that the integrated mass will reach unity at the lowest halo mass. Thus, unlike the HMF, the IMF naturally imposes a constraint on these fitting functions, which all need to obey. While the DMF predicts at which halo mass bin, the highest mass contribution is expected. These are detailed in \autoref{fig:idmf}.

\paragraph*{[IMF]}
The IMF is presented on the upper panels in \autoref{fig:idmf}. The first clear impression is that there seems relatively smaller scatter compared to the HMF, from around 0.6 to 2 at the lowest halo mass. That is expected because the upper bound of the IMF is 1, which means all the matter mass is within halos. Well, it is also clear that not all IMF derived from these HMF fitting functions approach unity eventually like PS. However, this feature of the PS analytical formula is by design. Some fitting functions cross the upper bound when the halo mass is around Earth's mass, while some plateau at different fractions until the smallest halo mass. Note that the IMF values at the smallest halo mass should be taken as a judgement of these fitting functions, though it would be great if this is taken into account for the newly fitting functions/parameters. 
For the FOF halos, besides the clear deviation from the PS result, all the other fitting functions seem to agree with each other within the shadow region. It is above about $10^{10} \hMsun$, where the total halo mass occupies half of the total universe mass. However, clear deviations already show up within the shadow region for the SO halos, even with the same overdensity parameter. Therefore, there is a wild spread out of the halo mass, at which half of the total universe mass is achieved. The bottom panels of the IMF plots show the relative differences to the PS result, which are basically in agreement with the upper panels.  

\paragraph*{[DMF]}
The DMF is shown on the lower panels in \autoref{fig:idmf}. As expected, there is a peak in the DMF due to the combination of halo mass and numbers. To indicate the peak position, a vertical thin dashed line at $M_{halo} = 10^{13} \hMsun$ is included in these plots. For the FOF halos, in agreement with the IMF, all the curves peak at the same position -- slightly larger $10^{13} \hMsun$, including PS with a different height. The contribution decreases on both high and low halo masses, albeit the high mass end decreases precipice-like, mostly due to the sharp drop in the HMF, compared to the low mass end. On the right-hand side, the SO halos again show large variations. However, these variations are shown mostly in the amplitude. While the peak positions are in very good agreement between different results, lying around the vertical think dashed line. This invariant peak position implies this can be used to constrain cosmology parameters, which will be detailed in the next subsection.

\subsection{The effects of cosmology parameters} \label{subsec:cosmo}

As mentioned in \autoref{sec:hmf}, the cosmology is fixed to the Planck 2015 cosmology parameters in the previous section. It would be interesting to know how the different cosmology parameters will affect these halo functions. However, before that, it is necessary to know whether the HMF is universal or not. If it is, this means these fitting functions can be directly used to predict the halo abundance in different cosmologies, with which only the input matter power spectrum is changed according to these parameters. If not, the cosmology dependence needs to be taken into account in these fitting functions, which makes the investigation of the cosmology parameter influence in this subsection impossible. Fortunately, \cite{Bocquet2020} investigated the HMF universal approach for cosmology model with over 100 different realisations and showed the accuracy is at about 10 per cent level up to about $10^{14} \hMsun$ and only worsens somewhat toward higher masses compared to an emulator. As such, it is safe to investigate how the different HMFs behave according to different cosmology choices by simply varying these cosmology parameters in \textsc{hmf}.

\subsubsection{The effects of $\sigma_8$}
\begin{figure}
    \centering
    \includegraphics[width=0.5\textwidth]{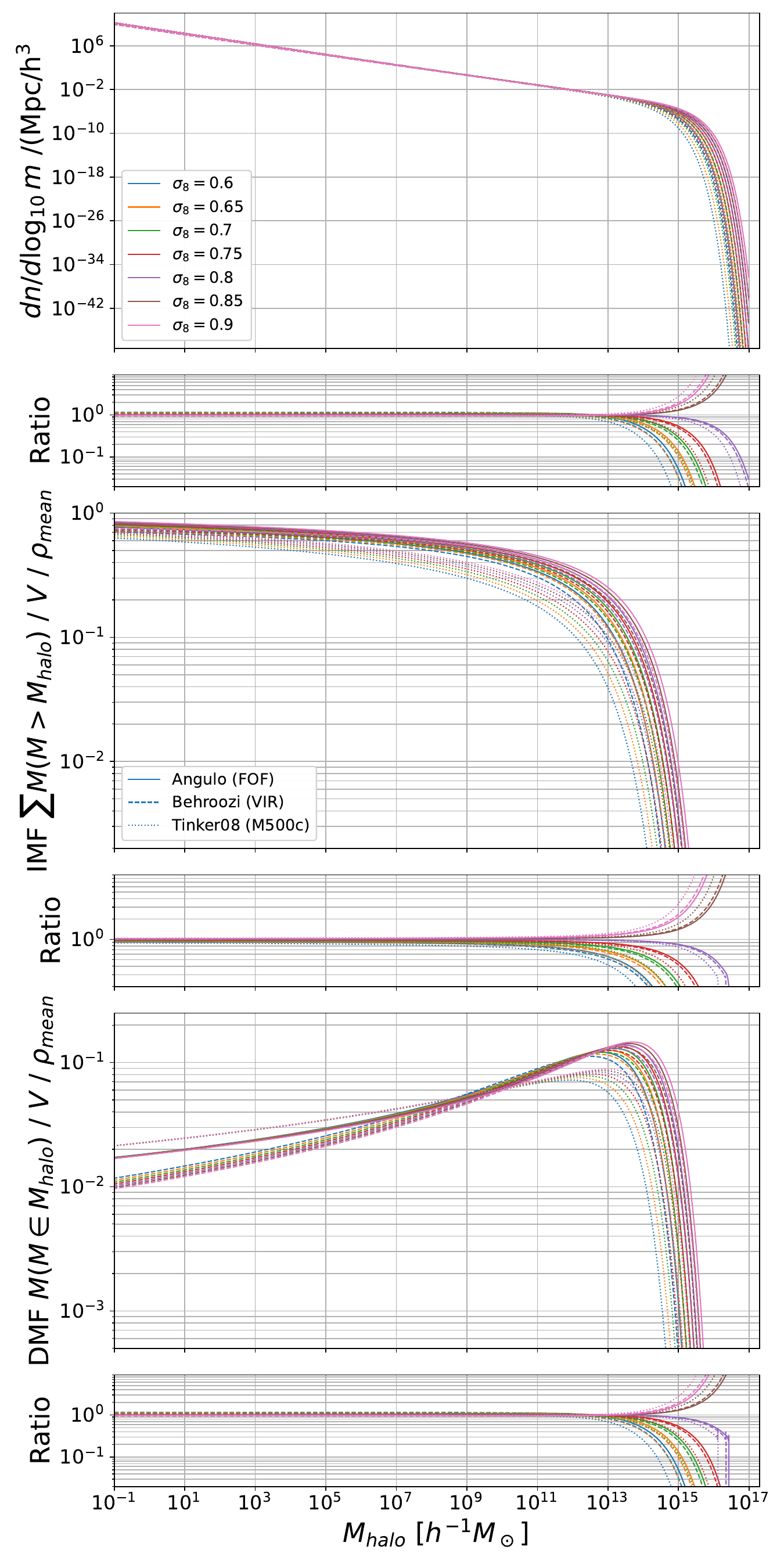}
    \caption{From top to bottom, the HMF, IMF and DMF are shown respectively. Three example fitting results from \citet{Angulo2012,Behroozi2013,Tinker2008} are shown in solid, dashed and dotted lines, respectively. Different colours, as indicated in the legend on the top panel, are for different values of the cosmology parameters -- $\sigma_8$ in this plot. These panels under each plot show the relative difference to the default Planck 2015 cosmology results of each fitting function. To highlight the result in the most interesting regions, the halo mass range is limited to $\gsim 10^{-1} \hMsun$. }
    \label{fig:sig8}
\end{figure}

$\sigma_8$, the mass variance on a scale of $8 \hMpc$, not only links to the power spectrum amplitude but also goes directly into HMF. As such, one can expect that it will leave a clear influence on these halo functions. Its influences are shown in \autoref{fig:sig8}. 

To my surprise, the change of $\sigma_8$ only affects these halo functions with $M_{halo} \gsim 10^{13} \hMsun$. It is clear that the larger $\sigma_8$, the higher HMF at the massive halo mass end. And the further away from the default $\sigma_8$, the affected halo mass can be lower. As indicated in the under panel, the relative difference is almost negligible for these low-mass halos. For the IMF, $\sigma_8$ leaves a slightly larger influence, mostly on the halo mass range. This is because it is an integration quantity, as such the more massive halos yield a consistent effect, though becomes less important at the low halo mass end. The DMFs from Angulo and Thinker08 are basically in line with their HMFs, i.e. only the massive halo mass end is affected. However, Behroozi, though no clear difference from the other two on HMFs, presents an opposite trend at the low halo mass end, i.e. the lower $\sigma_8$, the higher DMF fraction. Note that using SMT yields the same trend. Therefore, this could be because of the intrinsic definition of the VIR, which results in a junction point at $M_{halo} \approx 10^{10} \hMsun$. At a lower halo mass, the DMF is levered up with a lower $\sigma_8$. Furthermore, by comparing these three fitting functions in detail, Thinker08 tends to be influenced by the change of $\sigma_8$ mostly for all the halo functions at the massive end. This could be because the 500c tracks the highest-density region, which is highly sensitive to the change in power spectrum normalisation. Lastly, the peak position of DMF shifts towards the low halo mass as $\sigma_8$ decreases.

\subsubsection{The effects of $\Omega_m$}
\begin{figure}
    \centering
    \includegraphics[width=0.5\textwidth]{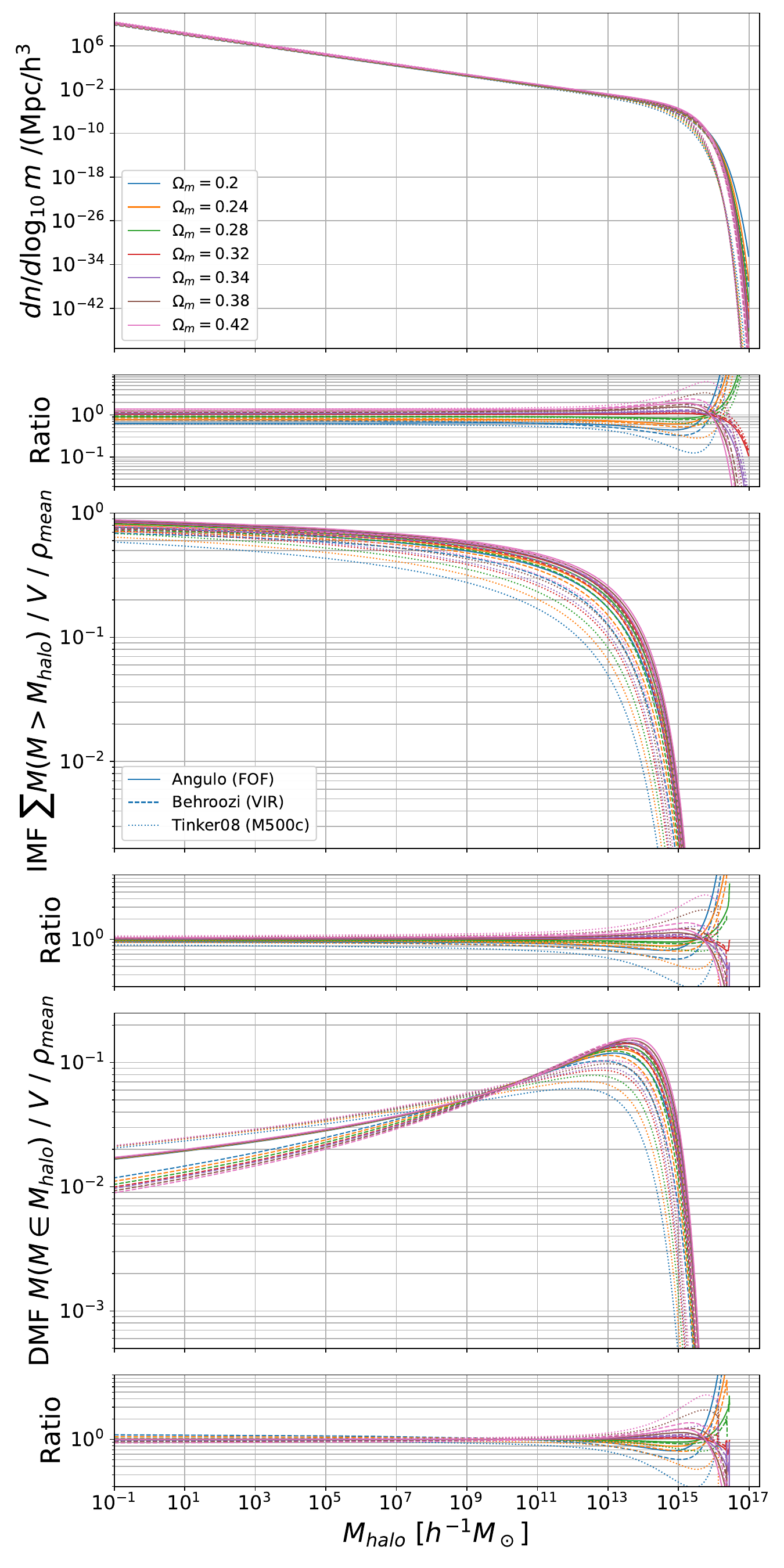}
    \caption{Similar to \autoref{fig:sig8}, but for different $\Omega_m$}
    \label{fig:om0}
\end{figure}

Similar to \autoref{fig:sig8}, the halo functions resulting from the different values of $\Omega_m$ are shown in \autoref{fig:om0}. Similar to the effect of $\sigma_8$, the halo functions at the most massive end still have the highest affection by $\Omega_m$. However, there are noticeable changes at the low halo mass range as well for all three halo functions. Again, it seems that these functions from Tinker08 with the 500c overdensity tend to be mostly affected by $\Omega_m$ at all halo mass ranges. One noticeable thing, different to $\sigma_8$ at the high-halo mass end, is that the decrease (increase) of $\Omega_m$ along with lower (higher) halo functions will turn to the opposite direction at the most massive end, at $M_{halo} \approx 10^{16} \hMsun$. This yields a feature, which can be used to break the degeneracy between $\Omega_m$ and $\sigma_8$. Lastly, the peak position in the DMF is also shifting towards low halo mass with smaller $\Omega_m$. 

\subsubsection{The effects of $n$}
\begin{figure}
    \centering
    \includegraphics[width=0.5\textwidth]{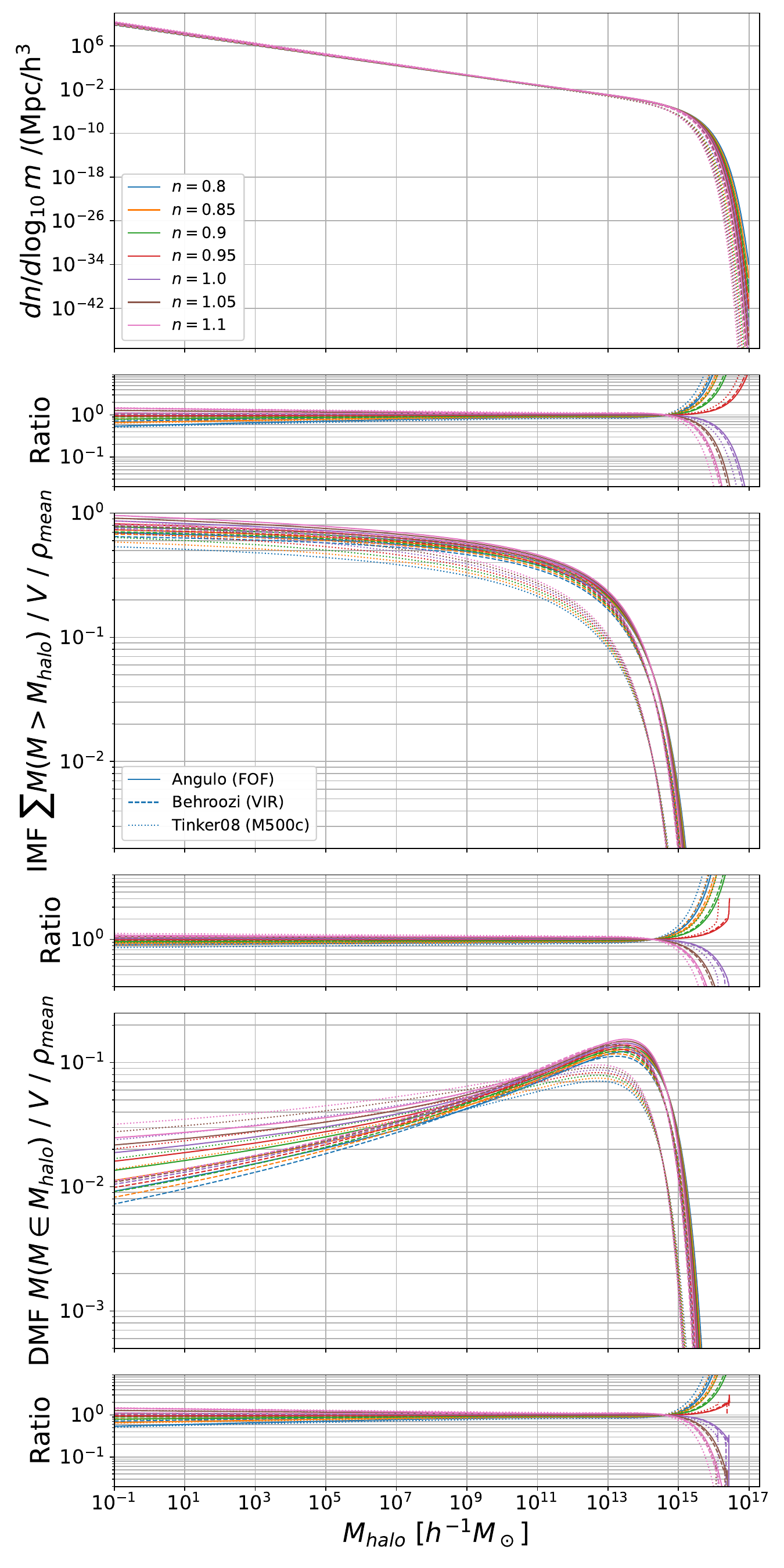}
    \caption{Similar to \autoref{fig:sig8}, but for different spectral indexes, n. }
    \label{fig:sin}
\end{figure}

As indicated in \autoref{fig:sin}, the spectral index, $n$, shows different influences on these halo functions due to the different values of $\sigma_8$: (1) both the massive and low mass halos are affected by $n$, albeit that the low mass halos still show relatively small changes compared to the most massive end and the massive halo mass range affected by $n$ is much larger, with a clear junction point at $M_{halo} \approx 10^{14.5} \hMsun$; The higher $n$ results in a lower halo function at the massive halo mass end and a opposite trend at the lower halo mass range; The peak position of the DMF is almost unaffected by the $n$.

\bigskip

The Hubble parameter is also investigated, however, its influence is very similar to other parameters not much different to the spectral index $n$ result at the massive halo mass end and to the the $\sigma_8$ results at the low halo mass end. More details can be found in \autoref{app:2}. 

\section{Conclusion and discussions} \label{sec:conc}

This paper is built on two assumptions: the halo's density profile is unchanged over all halo masses, and the HMF can be consistently described over all halo masses with one function. Both assumptions are confirmed or supported by the VVV project \citep[see ][respectively]{Wang2020, Zheng2024}. Based on this, the fitting functions for halo mass are extended down to dark matter particle mass and examined using the \textsc{hmf} package, which covers about 80 orders of magnitude in halo mass instead of 20 in the VVV project. Almost all fitting functions (except Jenkins) show good agreements with each other down to $\sim 10^{-60} \hMsun$, which corresponds to the dark matter particle mass at the GeV scale. Let's assume a simulation with infinite resolution, and its HMF should roughly follow the ones investigated in this paper with an extension. Therefore, one can expect that there will be about $10^{15}$ dark matter halos within one $\hMpc^3$ volume, i.e. about $10^{-3}$ dark matter halos between Earth and the Sun, or about hundreds of them within our solar system! If this rough estimation is true, maybe one day, we can directly detect these Earth-mass halos if they are exits. However, this is a normalised result, our solar system is located in a highly dense region of the Universe, where the exact number of Earth-mass halos may differ. In any case, it is fun to think about how the dark matter distributes in the local Universe, or even within our Milky Way galaxy.
Some interesting detailed findings are summarised here:
\begin{itemize}
    \item After extending different fitting functions down to dark matter particle mass. the variation of the dark matter halo number at the lowest mass is roughly within 2 orders of magnitude with the halo mass range being about 80 orders of magnitude and the HMF covering over 100. And this scatter doesn't depend on the halo definition or halo finding parameters.
    \item By proposing the normalised IMF and DMF, which describe the distribution of halo in integrated and differential masses respectively, the variation shows up more clearly. According to the IMF, half of the total universe's mass is achieved at different halo masses depending on the fitting functions. However, the peak position in DMF seems invariant and always at around $10^{13} \hMsun$, galaxy groups, for the given Planck 2015 cosmology model. This implies galaxy groups occupy the highest mass fraction in the universe.
    \item The influence of the cosmology parameters is investigated in \autoref{subsec:cosmo}. These halo functions at the end of the most massive halo have the strongest change due to the changes in these cosmology parameters. Both $\Omega_m$ and spectral index, $n$, can affect the halo functions at the low mass end.
\end{itemize}

Several things can affect the assumptions in this paper. One clear thing is the free-streaming scale of the dark matter particles, which is considered to be extremely small besides the two previous assumptions. Not to mention the warm dark matter, both of which can determine the smallest halo mass. If either is large, the conclusions of this experiment will be limited to the smallest halo mass. While above the halo mass limit, I believe the conclusion should stand. The other big uncertainty of this work is the influences from baryons, especially at high redshift for the formation of mini halos. It is still unclear whether these baryon processes can affect the formation of these tiny halos and how. Using high-resolution hydrodynamic simulation with ionization and heating of the gas, \cite{Chan2024} found that these mini-haloes can be photoevaporated, which will remove the gas in halos and lower the total mass of the halo. Furthermore, it has been shown that the baryons can affect the HMF, even at the massive halos mass end and at $z=0$ \citep[see][for example]{Cui2012,Cui2014}. 

In addition, different dark matter candidates will have different influences on these halo functions. For example, primordial black holes can also be dark matter, or at least partially substitute dark matter particles. \cite{Zhang2024} showed a bimodal feature in the HMF due to different fractions of dark matter replaced by primordial black holes. 

Though there are many many low mass halos as mentioned before, where to find them? Do they tend to live in high-density regions, such as a galaxy or low-density regions? By comparing the VVV project result with these fitting functions, it seems that their HMF from the void region only occupies a fixed fraction ($\sim 30$ per cent) of the total halo number. However, by assigning the whole simulation volume into 4 different environments \citep[e.g.][]{Cui2018,Libeskind2018,Cui2019}: knots, filaments, sheets and voids, \cite{Cautun2014} found that about 60 per cent of halos at $M_{halo} \approx 10^{10}\hMsun$ are in the void and sheet environment, with a clear increasing trend as halo mass decreases. The disagreement between the two could be because of the volumes of VVV simulations, which become extremely small at their higher level/resolution runs. Therefore, their results seriously suffer from the cosmic variance problem.

\section*{Acknowledgements}
The author gratefully thanks Steven Murray for (1) the publicly available {\textsc{hmf}} package; (2) his help; (3) the inspiration of his paper: How well do we know the halo mass function? The author would like to thank Alexander Knebe for valuable discussions. The author would also like to mention that the VVV project inspires this work. 

The analysis and plots for this study heavily rely on these Python packages: Ipython with its Jupyter notebook \citep{ipython}, astropy \citep{astropy:2018,astropy:2013}, NumPy \citep{NumPy}, SciPy \citep{Scipya,Scipyb}.and the matplotlib package \citep{Matplotlib}. 
Furthermore, this draft is written with the help of the \hyperlink{https://chromewebstore.google.com/detail/search-and-cite/opadkkneiklbpkpdglojbcgdfaanmkoh?utm_source=ext_app_menu}{\textsc{search and cite} extension} for fast literature research and citation adding.

WC is supported by the Atracci\'{o}n de Talento Contract no. 2020-T1/TIC-19882 was granted by the Comunidad de Madrid in Spain, and the science research grants were from the China Manned Space Project. He also thanks the Ministerio de Ciencia e Innovación (Spain) for financial support under Project grant PID2021-122603NB-C21 and HORIZON EUROPE Marie Sklodowska-Curie Actions for supporting the LACEGAL-III project with grant number 101086388. 

\section*{Data Availability}

The analysis and plotting scripts are publicly available from the Jupyter notebooks in this repository: \url{https://github.com/weiguangcui/HmkHMF}




\bibliographystyle{mnras}
\bibliography{example} 




\appendix

\section{The effect of transfer function}\label{app:1}
\begin{figure}
    \centering
    \includegraphics[width=0.5\textwidth]{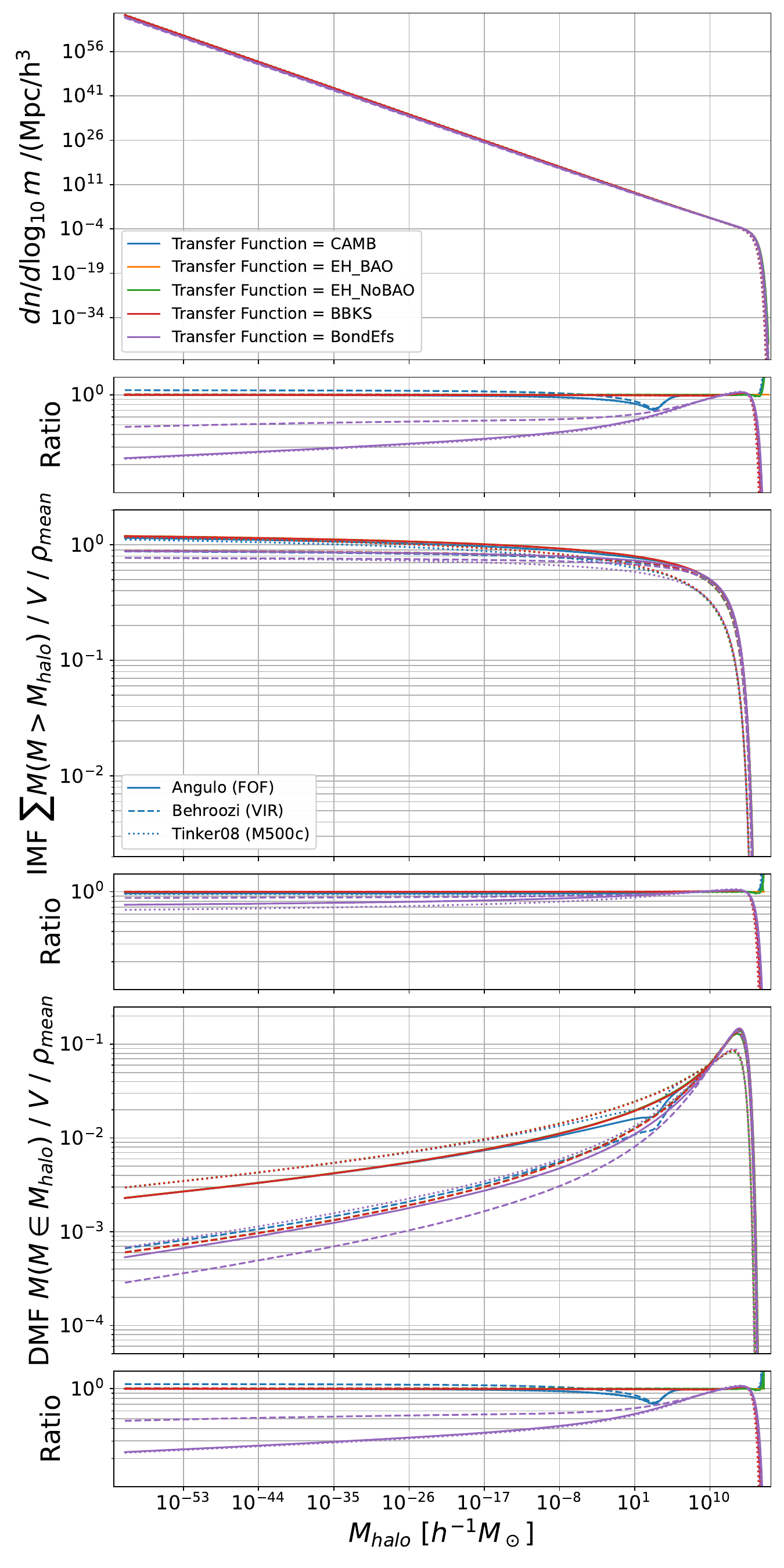}
    \caption{Similar to \autoref{fig:sig8}, but for the effects of different transfer functions. Note here that the same cosmology is assumed. And the relative difference is compared to the EH\_BAO \citep{Eisenstein1998} transfer function result. Furthermore, the halo mass is also extended to $10^{-60} \hMsun$.}
    \label{fig:trfu}
\end{figure}
Besides, the cosmology parameters, the transfer function may also influence these halo functions. As such, the effect of different transfer functions, provided inside the \textsc{hmf}, is probed in \autoref{fig:trfu}. There are 5 different transfer functions: CAMB \citep[with its latest Python package][]{Lewis:1999bs, Lewis:2002ah}; EH\_BAO and EH\_NoBAO \citep[with and w/o BAO wiggles][]{Eisenstein1998}; BBKS \citep{Bardeen1986}; BondEfs \citep{Bond1984}. As shown on the top panel in \autoref{fig:trfu}, there is very little change between these transfer functions, besides BondEfs, which decreases gradually compared to EH\_BAO as halo mass drops. A similar trend is also found for the IMF and DMF. However, there is a weird deep in CAMB at around $10^3 \hMsun$ compared to EH\_BAO.  

\section{The effect of Hubble parameter} \label{app:2}
\begin{figure}
    \centering
    \includegraphics[width=0.5\textwidth]{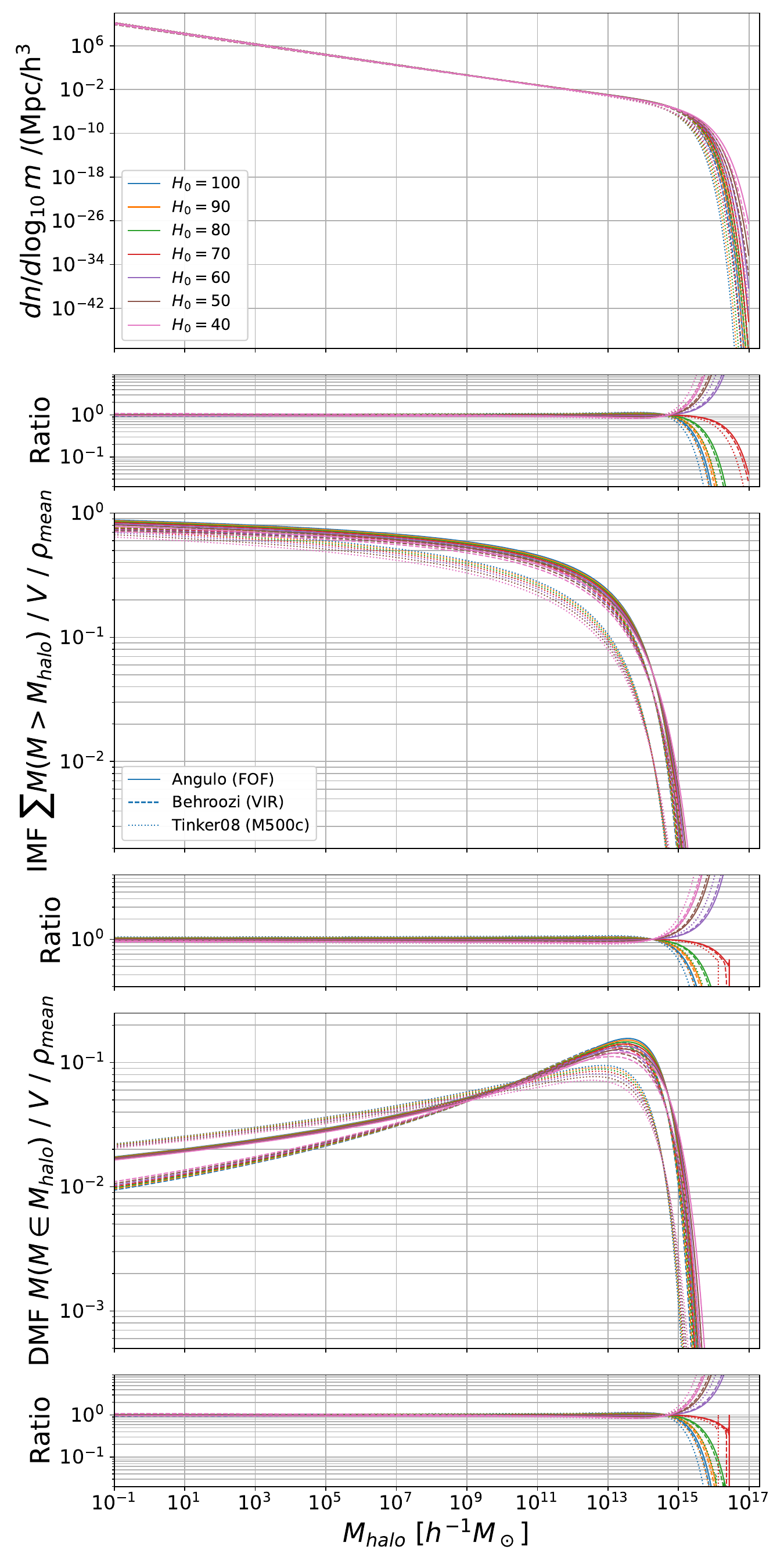}
    \caption{Similar to \autoref{fig:sig8}, but for the Hubble parameter at $z=0$.}
    \label{fig:hp0}
\end{figure}

A variety of Hubble parameters, as indicated in the legend on the top panel, are adopted for these halo functions. For the HMF, it is clear that only these most massive halos, $M_{halo} \gsim 10^{15} \hMsun$, are affected by it. The IMF roughly follows the HMF result but with a slightly larger change. The DMF also show weak dependence on the Hubble parameter at a low halo mass range, but its peak position seems unchanged. 


\bsp	
\label{lastpage}
\end{document}